# Reply to
# « A Comment on "The Far Future of Exoplanet Direct Characterization" - the Case for Interstellar Space Probes » by I. Crawford

Jean Schneider
Paris Observatory

**Introduction**

To reiterate the context, as mentioned by Fridlund and Lammer (2010), this paper resulted from the collective work of TE-SAT, a team appointed by ESA in 2003 to assess a strategy to find and characterize Terrestrial Exoplanets. The interstellar flight aspect was not part of the TE-SAT work and was added afterward as side remarks in the chapter on « The Far Future of Exoplanet Direct Characterization ».

As an introductory general remark, the intention of the paper was not to discourage work on interstellar flight prospective. On the contrary, any advance in this field that makes interstellar travel closer to us is welcome, and the author is gratefull to I. Crawford to provide an educated update and several recent references. The purpose of our paper was only to stress that, after the detection of first candidate biosignatures foreseeable in the coming 5-30 years, there will be a very long (frustrating) period during which we will have no clues as to the real size and morphology of biological organisms. My personal estimate is that that period may last centuries. Let me now reply in more detail to the Comment by I. Crawford. In the framework of this short reply, the discussion can only be qualitative; it would deserve a full future paper.

**Interstellar travel issues**

*Travel speed and propulsion*

As correctly mentioned by I. Crawford, the majority of papers do not use a travel speed of 0.3c, but rather 0.1c. The value of 0.3c of our paper was just a flag beyond which relativistic effects must be taken into account. Therefore, in this Reply I will use 0.1c.

It is not the place in this short Reply to discuss all propulsion mechanisms listed by I. Crawford ; I will only briefly comment on two of them—nuclear fusion and antimatter-powered rockets—as an illustration of practical difficultie.

The Comment puts its main hope for future propulsion on nuclear fusion. It is indeed conceptually simple, resting on solid physical concepts. But the road from concepts to realization may be very long. It is a fact that 70 years after the invention of a concept as simple as A+B-->C+ energy (D+T--> He+n +17.6 Mev in the present reactor projects), it has still not been possible to make it work at a stable energy production level. The ITER fusion facility is not expected to achieve production energy at a demonstration level before 2030, that is, almost a century after the nuclear fusion concept was invented. The author correctly mentions the developments in miniaturization. As an example, he cites the National Ignition Facility (a similar, less advanced project called « Mega Joule Laser » exists in Europe). But this facility, with all its control and cooling systems, is presently quite a non-miniaturized building. In spite of the fact that presenly it will only provide impulsive (non continuous) fusion energy, presently at a slow rate of one impulse per hour, one can imagine that in the future these impulses can be accumulated to provide a sufficient acceleration to the spacecraft. But it requires an initial energy of a few mega joules per 1 nanosecond impulse, and in the spacecraft this energy must come from somewhere.

For antimatter-powered rockets, there is a problem with production of antimatter (not speaking of onboard storage). The relative cross-section of anti-nucleon $N^-$ production in hadronic reactions like A+B--->C+N+$N^-$ is small. At 70 Gev incident energy, 1% of hadrons secondaries are $NN^-$ pairs (e.g., Tonwar et al 1971). The energy required to bring a 100 ton spacecraft at 0.1c is $10^{27}$ erg. The annihilation of one $NN^-$ pair provides $10^{-3}$ erg. It is thus necessary to embark, and first produce, $10^{30}$ $NN^-$ pairs. With a 1% efficiency of $NN^-$ pair production per nuclear reaction at 70 Gev, the total energy required to produce $0^{30}$ $NN^-$ pairs is finally 7 $10^{23}$ J or 200 terawatt during 10 years of continuous production. The present total instantaneous energy poduction on Earth is about 20 terawatt.

*Damage by interstellar dust.*

Let's take the numbers given by I. Crawford for interstellar dust density. The relevant quantity is the number $N$ of dust particles encountered by the spacecraft with a transverse section $S$ along a journey of length $L$. $N$ is given by

$$N = L.S.nu = L.S.rho/m$$

where $nu = rho/m$ is the number density of dust, $rho$ the mass density for dust grains of mass $m$. Let us take $L = 1pc$. For graines of mass $m = 10^{-12}$ kg, Crawford gives $rho = 6 \times 10^{-24}$ kg m$^{-3}$ leading to N = $10^8$ for a generously small transverse section $S$ = 10m x 10m. For 100-µm grains, Crawford gives $nu$ = 4

$10^{-17}$ m$^{-3}$, leading to $N = 400$. Crawford finds $N = 2$ per square meter, compatible with $N = 400$ for $S = $ 10m x 10m.

The point here is that, if 100-µm grains are dangerous, even two lethal or severe collisions are prohibitive. For $N ≤ 1$, the probability of 1 collision during the journey is $N$. The question is what probability of collision is acceptable. If a collision is lethal, this probability must be extremely close to zero for a several hunbred billion € mission.

*Remarks on the implementation of an interstellar travel mission*

The real implementation of an interstellar travel mission will necessarily follow the standard cost and risk evaluation criteria of Space Agencies. As a reminder, the cost increases when the risk is decreased. For such a complicated and costly mission as interstellar travel, the risk will have to be reduced well below the usual ~1% to 0.1% chances of failure as is the case for current missions, say down to 1 over several thousand, which will therefore increase the cost. The risk evaluation is traditionally based on « Technology Readiness Level » (TRL). TRLs are categorized, according, for instance, to ESA's Science Technology Development Route (ESA 2010) in 9 levels from 1 (Basic Physical Principles) to 9 (Actual system "Flight proven" through successful mission operations). For interstellar travel, we are at best at level 1 (or even 0.5), while a « Flight proven » mission will realistically require first a precursor mission to secure the technological concept, including shielding mechanisms, at say 500 to 1000 Astronomical Units. As a comparison, I can take the nulling interferometry concept for the infrared detection of exo-Earths. It was invented in the late 70s (Bracewell 1978) and is still not forseen for a launch by ESA and NASA before 2030, that is, 50 years after the invention of the concept for a mission at least 100 times easier and cheaper than interstellar travel. It is true that, as an opposite example, the energy reached by particle accelerators (from first cyclotrons to LHC) has been multiplied by a factor 1 million in about one century. But in space the probability of failure must be very close to 0.

A further issue is interstellar communications. The problem is to reconcile a small mission with the high power needed to send signals from the spacecraft at several light years down to Earth. The spot of a signal received on Earth from a 100 m wide antenna onboard a spacecraft at 4 light years (the alpha Cen distance) at optical (laser) wavelength is of the order of 1000 km; detecting this signal would require a very large antenna on Earth or a very powerfull laser onboard, or both. I take this opportunity to propose another way to detect signals from the spacecraft. It would consist on a « Venetian shade » -like setup onboard the spacecraft where the signal is coded by a series of successively open and closed shades in the line of sight of the target star as seen from Earth. The resulting temporal variations of the diming of the stellar flux would be observed at very high angular resolution with a hypertelescope (Labeyrie 1996, Bouyeron et al. 2010). Of course, maintaining the spacecraft in the line of sight of the star would require very precise navigation.

Another very crucial aspect is the choice of the target. To deserve an interstellar travel mission, an exoplanet will require very solid clues of biosignatures (to quote Willy Benz, « Extraordinary claims require exceptional proofs »). I hope that current radial velocity monitorings will discover the existence of habitable planets around Alpha Cen A or B, and that in the coming decades these planets will reveal solid biosignatures. But what if the nearest planet with credible biosignatures lies at 10 pc? Even at a speed of 0.1c, the travel will last 400 years.

**Conclusion**

It is presomptuous to predict exactly what will happen after one century and into the future, but it is more than likely that development of the capacity to observe the morphology of meter-sized organisms on exoplanets will take several centuries, at least in the framework of present and forseable physical concepts. Another optimistic possibility would be that, in a nearer future, we will detect pictures of extraterrestrials with a good resolution in SETI signals. The debate must still go on.

**References**
Bouyeron L., Olivier S., Delage L. et al. 2010. MNRAS, Accepted. arxiv:1009.1953
Bracewell R., 1978. Nature, 274, 780
ESA's Advanced Studies and Technology Preparation Division, 2010. *Strategic Readiness Level.* Available at
http://sci.esa.int/science-e/www/object/index.cfm?fobjectid=37710
Fridlund M. and Lammer H. 2010. *Astrobiology*, 10, 1
Labeyrie A., 1996. Astron. & Astrophys. Suppl. 118, 517
Tonwar S., Naranan S. and Sreekantan B., 1971. Lett. Nuov. Cim. 1, 531